\documentclass[prl,aps,reprint,showpacs,superscriptaddress]{revtex4-1}
\usepackage{amsmath,amssymb}
\usepackage{graphicx}

\begin{document}

\title{Branching rate expansion around annihilating random walks}

\author{Federico Benitez}
\affiliation{LPTMC, CNRS-UMR 7600, Universit\'e Pierre et Marie Curie, 75252 Paris, France}
\affiliation{Instituto de F\'{\i}sica, Facultad de Ciencias, Universidad de la Rep\'ublica, 11400 Montevideo, Uruguay}
\author{Nicol\'as Wschebor}
\affiliation{Instituto de F\'{\i}sica, Facultad de Ingenier\'{\i}a, 
Universidad de la Rep\'ublica, 11000 Montevideo, Uruguay}

\begin{abstract}
We present some exact results for Branching and Annihilating Random Walks. We compute the non-universal threshold value of the annihilation rate for having a phase transition in the simplest reaction-diffusion system belonging to the Directed Percolation universality class. Also, we show that the accepted scenario for the appearance of a phase transition in the Parity Conserving universality class must be improved. In order to obtain these results we perform an expansion in the branching rate around Pure Annihilation, a theory without branching. This expansion is possible because we manage to solve Pure Annihilation exactly in any dimension.
\end{abstract}

\pacs{05.10.Cc 64.60.De 64.60.ae 82.20.-w} % end of PACS codes

\date{\today}% It is always \today, today,
\maketitle

% The study of out of equilibrium systems stands out as one of the most challenging problems in statistical mechanics. Generically, these systems exhibit a much richer variety of behavior than its equilibrium counterparts. This is specially true when regarding phase transitions and critical phenomena, where most established methods fail to deliver accurate results. 
%, with the absorbing phase corresponding to a no-fluctuations state,

Active-to-absorbing phase transitions represent one of the simplest cases where genuine non equilibrium behavior is attained. In this context, much work has been devoted to the study of Branching and Annihilating Random Walks (BARW) \cite{henkel,grassberger84,cardy96,tauber05,janstauber}, systems composed of particles of a single species $A$, that diffuse in a $d$-dimensional space, and that can suffer both annihilation and branching (i.e. offspring creation) processes. BARW are not only of direct physical interest, but also present a conceptually simple class of out of equilibrium systems. 

Due to universality, it is generally enough to consider the simplest possible reactions, which allow for BARW to be classified into sub-classes \cite{henkel,cardy96}. Pure Annihilation (PA), the theory without branching where the only reaction is $2A\xrightarrow{\lambda} \emptyset$, constitutes a good starting point in order to study properties of BARW. In the long time limit the PA system approaches the empty state, where all density correlation functions are zero. The response functions, however, are non trivial, and are governed in the Infrared (IR) (that is, for momenta and frequencies smaller than the scale set by $\lambda$) by a non-Gaussian RG fixed point. If we add the simplest branching reaction  $A \xrightarrow{\sigma} 2A$, the resulting BARW system is known to be in the Directed Percolation universality class \cite{janssen81}, and we call it BARW-DP. If instead we choose to add the reaction $A \xrightarrow{\sigma} 3A$, which preserves the parity of the number of particles, the system belongs to 
the Parity Conserving (PC) universality class (more appropriately known as Generalized Voter \cite{chate}) whenever a phase transition takes place. We call this system BARW-PC \cite{notepcpd}.

% The addition of a branching reaction to PA has been studied using perturbative RG methods \cite{cardy96}, by expanding the model around the Gaussian reaction-less RG fixed point. There, the authors show that the PA fixed point becomes unstable for all dimensions when the reaction  $A \xrightarrow{\sigma} 2A$ is included, and becomes unstable for all $d>d_c$ for $A \xrightarrow{\sigma} 3A$, with $d_c$ generally believed to lie between one and two. This implies the existence of a new fixed point $F^{PC}$ for $d<d_c$, which would govern the known $d=1$ PC phase transition.

Both universality classes have been studied using perturbative RG methods \cite{cardy96}, by expanding the model around the Gaussian reaction-less RG fixed point. In the case of DP, the authors of \cite{cardy96} do not find a phase transition in dimensions $d>2$. As for PC, at 1- and 2-loop orders an (upper) critical dimension $d_c>1$ is found, with the PA fixed point being unstable for $d>d_c$, so that the branching perturbation is always relevant, there is no absorbing phase, and hence no phase transition. Conversely, for $d<d_c$ the PA fixed point is fully attractive and an absorbing phase exists at small branching $\sigma$, whereas at larger $\sigma$ the system is in its active phase. A phase transition must therefore occur at finite $\sigma$, which is confirmed numerically in $d=1$. In \cite{vernon}, a 1-loop analysis of BARW with L\'evy-flight dynamics was shown to be compatible with this scenario. Finally, Non Perturbative Renormalization Group (NPRG) studies are also consistent with it  \cite{
canet3}.

In this letter, we revisit these conclusions, proving that: (i) in DP there is a phase transition for all $d$, in agreement with what is
found in Montecarlo and NPRG studies \cite{canet1,canet2}, and (ii) in PC, the PA fixed point remains unstable in the $\sigma$-direction for all relevant dimensions $d \geq 1$, and therefore there are aspects of the PC transition that are still to be fully understood.

Our results rely on an expansion in $\sigma$ around the PA theory (contrary to the usual perturbative expansion, which is performed around the Gaussian theory). This expansion is highly non trivial since it requires to solve exactly PA, which we do by deriving closed equations for all its response functions. As far as we know, such an expansion around a non-Gaussian model has never been performed before for out of equilibrium systems. Since our approach is valid for any value of the annihilation rate $\lambda$, we obtain exact results at small $\sigma$.  

Being based on an exact solution of PA, our approach also allows us to exactly compute non-universal quantities. We choose to calculate, for the BARW-DP model, the non-universal threshold value $\lambda_{th}$ of the annihilation rate, at which a phase transition occurs at vanishing $\sigma$.

\paragraph{Pure Annihilation.} We briefly recall some technical aspects. In order to study reaction-diffusion processes, a field theory can be constructed in a standard way by using the Doi-Peliti formalism \cite{doi76b}. As a result, one obtains the generating functional of the correlation and response functions 
\begin{equation}\label{ZJ}
{\cal Z}[J,\hat J] = \int {\cal D}\phi {\cal D} \hat \phi \exp \left(-S[\phi,\hat \phi]+\int_{x} J \phi+ \hat J \hat \phi\right)
\end{equation}
with an appropriate action $S[\phi,\hat \phi]$, that captures the microscopic reactions. Here we have introduced the notation, to be used throughout, $x=(\mathbf{x},t)$ (and $p=(\mathbf{p},\nu)$ in Fourier space), $\int_x=\int  d^dx\, d t$ and $\int_p = \int \frac{d^d p}{(2\pi)^d} \frac{d\omega}{2\pi}$.

In the case of PA the only reaction is $2 A\xrightarrow{\lambda} \emptyset$ and
\begin{equation}
\label{S_PA}
S^{PA}[\phi, \hat \phi] = \int_x \Big(\hat \phi (\partial_t - D \mathbf{\nabla}^2) \phi + \lambda (\hat \phi^2 -1)\phi^2\Big)
\end{equation}
where $D$ is the diffusion constant and where we have ignored initial conditions, because we are only interested in the steady state. This theory has a strong resemblance with equilibrium $\phi^4$ theory, but is further constrained by causality properties \cite{henkel}.

All the information of a system is encoded in the vertices $\Gamma^{(n,m)}$ of the theory (also known as the 1-Particle Irreducible (1PI) functions, with $n$ incoming legs and $m$ outgoing response legs), related to the connected correlation and response functions by a Legendre transform \cite{ZINN}. We now present an identity allowing us to obtain a closed equation for any $\Gamma^{(n,m)}$ in PA. It can be most conveniently written at the diagrammatic level: any Feynman diagram contributing to $\Gamma^{(n,m)}$ that  includes at least one loop must begin by a 4-legs bare vertex.
%%%%%%%%%%%%%%%%%%
\begin{figure}[ht]
\includegraphics[width=0.45\textwidth]{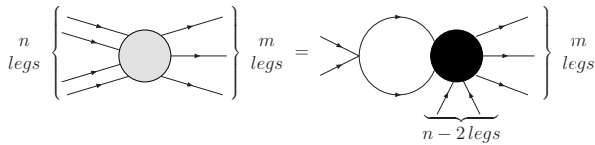}
\caption{Generic form of a diagram for $\Gamma^{(n,m)}$ including at least one loop. Left hand side: diagrammatic representation of a generic $\Gamma^{(n,m)}$ vertex. Right hand side: general structure for  $\Gamma^{(n,m)}$ in PA, the black blob is a connected and amputated Green function that has to comply with some requisites, see text.}
\label{gamnm}
\end{figure}
%%%%%%%%%%%%%%%%%%

In Fig. \ref{gamnm} we show the general structure of these diagrams. The black blob denotes a sub-diagram that is constrained by the condition that the full diagram must be 1PI. In particular, it means that this sub-diagram must be connected (and with amputated external legs). Now, any connected diagram contributing to this blob has a unique tree decomposition in terms of 1PI sub-diagrams having at most $n$ incoming and $m$ outgoing legs \cite{ZINN}. By summing all possible diagrams and permutations compatible with the 1PI structure of the full diagram, we obtain a closed equation that relates any $\Gamma^{(n,m)}$ with other $\Gamma^{(l,s)}$ with a lower number of legs. A non-perturbative proof (not based on an all-order analysis) of this general property will be given elsewhere \cite{largo}. 
%This same construction can be performed singularizing two outgoing legs, yielding equivalent results.

In order to be concrete, let us analyze the identity given in Fig. \ref{gamnm} for the simplest vertices. For $\Gamma^{(1,1)}$ this yields a well-known property: there is no correction to $\Gamma^{(1,1)}$ in PA, since there is no diagram such as the one in Fig. \ref{gamnm} with a single incoming leg. Concerning $\Gamma^{(2,2)}$, one arrives at
the known equation \cite{peliti,frey}
% %%%%%%%%%%%%%%%%%%
% \begin{center}
% \begin{figure}[ht]
% \includegraphics[width=0.45\textwidth]{BARW2.eps}
% \caption{Closed equation for $\Gamma^{(2,2)}$ in PA.}
% \label{gam22}
% \end{figure}
% \end{center}
% %%%%%%%%%%%%%%%%%%
\begin{multline}
\label{eqGamma22}
\Gamma^{(2,2)}(p_1,p_2,\bar p_1,\bar p_2)= 4 \lambda - 2 \lambda \int_q G(q) \\ 
\times G(p_1+p_2-q) \Gamma^{(2,2)}(q,p_1+p_2-q,\bar p_1,\bar p_2)
\end{multline}
with $G(q)=[\Gamma^{(1,1)}(-q)]^{-1}$ the propagator of the theory. The solution of (\ref{eqGamma22}) is of the form
\begin{equation}\label{deflambda}
\Gamma^{(2,2)}(p_1,p_2,\bar p_1,\bar p_2)= 4 l(p_1+p_2)
\end{equation}
with
\begin{equation}
\label{lambda(p)}
l(p)= \lambda\left(1+ 2 \lambda \int_q G(q) G(p-q)\right)^{-1}.
\end{equation}
This result can simply be seen as stemming from a geometric sum over bubbles. A similar relation is found for $\Gamma^{(2,1)}$. The relation presented in Fig. \ref{gamnm} is a generalization of these known results to any vertex function.
It enables us to study BARW by means of a perturbative expansion in the branching rate $\sigma$ around PA.

\paragraph{BARW-PC.} Let us begin by considering the Parity Conserving universality class: $2A \xrightarrow{\lambda} \emptyset$ and $A \xrightarrow{\sigma} 3A$. The corresponding microscopic action reads
\begin{equation}
 S^{PC}[\phi, \hat \phi] = \int_x \Big(\hat \phi (\partial_t - D \mathbf{\nabla}^2) \phi + \lambda (\hat \phi^2 -1)\phi^2 \\ + \sigma (1-\hat \phi^2) \phi \hat \phi \Big).\notag
\end{equation}
This action is symmetric under $\phi \to -\phi$, $\hat \phi \to -\hat \phi$, which implies the constraint $\Gamma^{(n,m)}=0$ if $(n+m)$ is odd.

The PC model is known to present an active-to-absorbing phase transition in $d=1$, generally believed to be related to a change of stability of the PA fixed point in a dimension $d_c$ with $2>d_c>1$. Perturbatively \cite{cardy96}, and also within the Local Potential Approximation (LPA) of the NPRG \cite{canet3}, this change of stability occurs in the following way: On the one hand, for dimensions close to two, canonical power counting arguments show that the PA fixed point is unstable in the $\sigma$-direction, which implies that the system is in its active phase for all $\sigma>0$. On the other hand, at 1- and 2-loop orders an (upper) critical dimension $d_c>1$ is found such that for $d<d_c$ the coupling $\sigma$ becomes irrelevant around the PA fixed point, which therefore becomes fully attractive, thus showing the existence of an absorbing phase at small $\sigma$. This change of stability occurs because a new fixed point $F^{PC}$ crosses the PA fixed point at $d_c$ and in this dimension they both change 
their stability. Below $d_c$, this new fixed point is in the physically relevant quadrant $\lambda \geq 0$, $\sigma \geq 0$, has one unstable direction, and is thus associated with the PC phase transition. We now reconsider this scenario.

We can perform a systematic expansion in $\sigma$ while keeping $\lambda$ finite. Within this formalism we can reanalyze the stability of the IR PA fixed point in the presence of the PC creation reaction, $A \xrightarrow{\sigma} 3A$, that we can determine exactly since our analysis is exact at small $\sigma$. The relevance of this coupling can be obtained from the flow of $\Gamma^{(1,3)}$, since this function is of order $\sigma$.
%%%%%%%%%%%%%%%%%%
\begin{figure}[ht]
\includegraphics[width=0.45\textwidth]{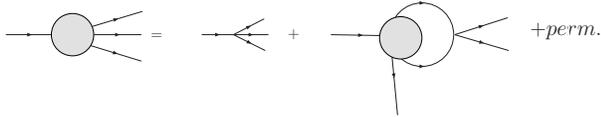}
\caption{Closed equation for $\Gamma^{(1,3)}$ in BARW-PC.}
\label{gam13even}
\end{figure}
%%%%%%%%%%%%%%%%%%

At first order in $\sigma$, we can write the equation for $\Gamma^{(1,3)}$ in the form shown diagrammatically in Fig \ref{gam13even}, whose structure implies the following functional form for $\Gamma^{(1,3)}$
\begin{equation}\label{defsigma}
 \Gamma^{(1,3)}(p,p_1, p_2, p_3)= -2\sigma(p, p_1)-2\sigma(p, p_2)-2\sigma(p, p_3).
\end{equation}
The quantity that interest us is $d_\sigma$, the scaling dimension of $\sigma$ in the IR limit, which can be extracted from the behavior of $ \hat\sigma( p)\equiv \sigma(0,-p)/l(p)$
\begin{equation}
 \hat \sigma(p)\sim |\mathbf{p}|^{d-d_\sigma}, \quad l(p)\sim |\mathbf{p}|^{2-d}  \quad \mathrm{for}\quad \nu,|\mathbf{p}|^2 \ll \lambda^{\frac{2}{2-d}}.\notag
\end{equation}
The equation for $\hat \sigma$ obtained from Fig \ref{gam13even} and Eq. (\ref{defsigma}) reads
\begin{equation}\label{sigmahat1}
 \hat \sigma(p)=\frac \sigma \lambda -4\int_q G(q) G(p - q) \hat \sigma(q) l(q).
\end{equation}
Using this exact expression and expanding in $\epsilon=2-d$ we easily recover the 1- and 2-loop results \cite{largo,cardy96}. 

In order to get a result for $d_\sigma$ it is convenient to get rid of the bare reaction rates, as we are only interested in the universal IR scaling behavior. We subtract to (\ref{sigmahat1}) its value at zero $\hat \sigma(p=0)$, which is zero for $d<2$, given that we expect $d_\sigma<d$. Our results will later confirm this. We must also take into account the expected scaling invariance. We define the scaling function $\tilde \sigma(\tilde \nu)$
\begin{align}\label{scaling}
 \hat \sigma(\mathbf{p},\nu)&=|\mathbf{p}|^{d-d_\sigma}\tilde \sigma (\tilde \nu), & \tilde \nu &=\frac{\nu}{|\mathbf{p}|^2}.
\end{align}
Observe that we are performing a perturbation around the PA fixed point, whose anomalous dimensions are zero, and the natural scaling variable is accordingly $\tilde \nu=\nu/|\mathbf{p}|^2$.

The ensuing equation for $\tilde \sigma$ is still too complicated to be solved analytically, and requires a numerical solution. A convenient way to do this is to make an expansion in $u=\cos \widehat{(\mathbf{p}, \mathbf{q})}$, which turns out to be rapidly convergent. We then proceed as follows: at each order in $u$, we adjust $d_\sigma$ at a given value of $d$, by numerically iterating this equation in order to reach a fixed functional form for $\tilde \sigma (\tilde \nu)$ in a lattice of $N_\nu$ points with a resolution $\delta \nu$. We have checked the convergence in $u$ and in the numerical parameters $\delta \nu$ and $N_\nu$, used for the computation of integrals. This  procedure gives always a converged scaling function $\tilde \sigma (\tilde \nu)$, which confirms a posteriori the scaling form ansatz (\ref{scaling}). We observe $\tilde \sigma (\tilde \nu)$ to be a non-trivial function of its argument \cite{largo}, which may explain the qualitative difference between our results and previous 
approximate results. Observe that LPA and 1-loop analysis are based on a $\Gamma^{(1,3)}$ vertex without dependence on frequency and momentum. 
%%%%%%%%%%%%%%%%%%
\begin{figure}[ht]
\includegraphics[angle=-90,width=0.45\textwidth]{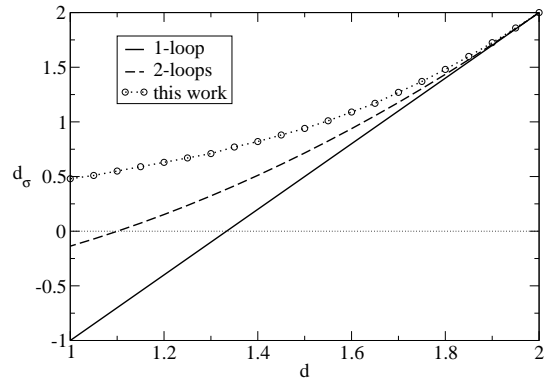}
\caption{Results for $d_\sigma$, showing there is no change in the RG relevance for the branching rate $\sigma$.}
\label{orden6}
\end{figure}
%%%%%%%%%%%%%%%%%%

This procedure allows us to find the value of $d_\sigma$ as a function of $d$, the results of which are plotted, together with previous perturbative results, in Fig. \ref{orden6}. There one sees that, even if $d_\sigma$ gets smaller when $d$ decreases, it remains always positive, which is in contradiction with the usual picture for the PC transition. Notice that this result does not rule out the possibility of a transition in $d=1$. We propose the following scenario that reconciles all existing results. In a dimension between $1$ and $2$ two fixed points appear at positive $\lambda$ and $\sigma$, the one with the smaller value of $\sigma$ being fully attractive and governing the absorbing phase, while the other is once unstable and is thus associated with the PC transition. In this scenario, the absorbing phase must have a behavior different from PA. This can be studied either by using Montecarlo methods or higher order NPRG equations.

The previous result is surprising because another exact analysis \cite{takayasu92} in $d=1$, which seems to indicate that $\sigma$ is irrelevant with respect to the
IR PA fixed point, in contradiction with our conclusions. We can explain this difference observing that the model used in \cite{takayasu92} is defined with $\lambda=\infty$ (and indeed presents no phase transition at all for whatever value of $\sigma$). Now, the IR limit of the theory corresponds to $\nu, |\mathbf{p}|^2 \ll \lambda^{2/(2-d)}$, but this does not allow us to take $\sigma=0$ when compared to $\lambda$. Looking at Eq. (\ref{sigmahat1}), $\lambda=\infty$ implies $\hat \sigma\equiv0$, so that the relevant direction corresponding to $\sigma$ is no longer accessible by studying $\sigma$ as a perturbation. This is true for all $d$. In particular, for $d\sim 2$, the form of the relevant direction can be calculated perturbativelly and the result is at 1-loop 
\begin{equation}
 \hat \sigma(p) \sim \frac {\sigma} {\lambda^3}  l^2(p)
\end{equation}
One observes then that in the limit $\lambda\to \infty$ the relevant direction is eliminated artificially even at $d\sim 2$. Thus, the results of \cite{takayasu92} do not apply to BARW-PC at finite values of the reaction rate $\lambda$, the system in which we are interested in. It also shows that when $\lambda$ is large, a cross-over must occur and for a long transcient the PA behaviour will show up. Montecarlo studies of the low branching regime of this system have until now, as far as we know, also been mostly made in the limit $\lambda\to \infty$ \cite{jensen,odorrev}. They are compatible with the standard scenario, but within the criticims previously pointed out.

\paragraph{BARW-DP.} Let us now consider the simplest BARW-DP model: $2A\xrightarrow{\lambda} \emptyset$ and $A\xrightarrow{\sigma} 2A$, whose microscopic action reads \cite{cardy96}
\begin{equation}
S^{DP}= \int_x \Big(\bar \phi (\partial_t - D \mathbf{\nabla}^2) \phi + \lambda \bar \phi (\bar \phi +2)\phi^2 - \sigma \bar \phi (\bar \phi + 1)\phi \Big). \notag
\end{equation}
Notice that in this equation we have performed, as is usually done, a shift in the response fields, $\hat \phi(x) = 1+ \bar \phi(x)$. The case $\sigma=0$ corresponds to PA, now written in terms of the shifted $\bar \phi$ field. This version of PA can be solved following the same ideas as previously.  In particular, it is easy to check that the equation for $\Gamma^{(2,2)}$ remains the same as in the unshifted case, Eq. (\ref{eqGamma22}). 

\begin{figure}[ht]
\includegraphics[width=0.45\textwidth]{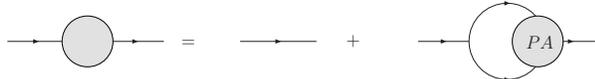}
\caption{Closed equation for $\Gamma^{(1,1)}$ in BARW-DP at $\mathcal{O}(\sigma)$.}
\label{gam11odd}
\end{figure}

%It is important to notice that this $\sigma$-expansion generates a \emph{convergent} series, something not very common when dealing with perturbative expansions in field theories. 
%This property follows from Lebesgue's dominated convergence theorem, given that we have under control the PA model (even in a  nonperturbative sense \cite{largo}).
We consider the exact calculation of the threshold $\lambda_{th}$ for the existence of an active-to-absorbing phase transition in BARW-DP. In order to check for the presence of a phase transition it is enough to study the behavior of $\Delta=\Gamma^{(1,1)}(p=0)$ as a function of the annihilation rate $\lambda$, since the zeroes of $\Delta$ correspond to a divergence of the correlation length. Given that $\lambda_{th}$ corresponds to the transition value of $\lambda$ when $\sigma\to 0^+$, an analysis at leading order in $\sigma$ allows for an exact calculation of $\lambda_{th}$. An equation for $\Gamma^{(1,1)}(p)$ at order $\mathcal{O}(\sigma)$ can be represented in the diagrammatic form of Fig. \ref{gam11odd}, which can be written
\begin{align}
\Gamma^{(1,1)}(p)& =-\sigma +4\sigma l(p) \int_{q} G(q)G(p-q)  + \mathcal{O}(\sigma^2)\label{flow_delta}
\end{align}
where we have evaluated the propagator $G(p)$ and the vertex $\Gamma^{(2,1)}(q,p-q,-p)$ at order zero in $\sigma$, that is in PA, and consequently replaced this last function by $4 l(p)$ \cite{cardy96,largo}. 
%%%%%%%%%%%%%%%%%%
%%%%%%%%%%%%%%%%%%!

By substituting the expression for $l(p=0)$, evaluating (\ref{flow_delta}) at $p=0$, and putting $\Delta=0$, we find
\begin{equation}
\label{lambdac}
 \lambda_{th}=\left(2 \int_q G(q)G(-q)\right)^{-1}.
\end{equation}
To evaluate $\lambda_{th}$, we need to take into account that the properties of a phase diagram are not universal, and depend on the specific form of the theory at short distances. This is as in equilibrium statistical mechanics, where critical temperatures depend on the specific form of the lattice. Here we consider a particular microscopic form for the model corresponding to a hypercubic lattice with lattice spacing $a$. In this case the propagator reads 
\begin{equation}
 G(q)=\frac{1}{i\omega+\frac{2D}{a^2}\sum_{i=1}^d (1-\cos(a q_i))}.
\end{equation}
In Table \ref{table}, the value of the resulting threshold coupling $\lambda_{th}$ is given, proving in particular the existence of a phase transition in every dimension. Previous results from Monte-Carlo simulations \cite{canet2,odor1} are in excellent agreement with these exact ones. This same general structure of the phase diagram has been also been show to exist in other models in the DP universality class \cite{odor2}.
\begin{table}[tp]
\begin{tabular}{lcccc}
\hline
  $d$                   &   3  &  4   &  5   &   6   \\ \hline
 $\lambda_{th}/D a^{d-2}$ (this work)  & 3.96 & 6.45 & 8.65 & 10.7 \\
 $\lambda_{th}/D a^{d-2}$ (MonteCarlo) \cite{canet2}  & 3.99  & 6.48 & 8.6 & 10.8 \\
\hline
\end{tabular}
\caption{\label{table} Values of the threshold coupling $\lambda_{th}$ for various dimensions $d$.}
\end{table}

It is convenient to point out that for $d\leq 2$ an IR divergence of the integral in (\ref{lambdac}) takes place. This makes $\lambda_{th}=0$ in those dimensions, in agreement with the results of \cite{cardy96}. For this reason, for $d\leq 2$ it is not useful to expand the model at small $\sigma$ for a finite $\lambda$ in order to study the phase transition. Moreover, this also shows that in those dimensions the transition is dominated by IR effects, and correspondingly most of the dependence on the microscopic behaviour of the model is absent.

\paragraph{Concluding remarks.} In this work we take advantage of the structure of PA to find closed exact expressions for any of its response functions, which we use to perform an expansion in the branching rate $\sigma$ around this model. This gives us access to the small branching regime of BARW in two important universality classes.

In the case of the system of reactions $2A\to \emptyset$, $A\to 2A$, which belongs to the DP universality class, we give an explicit proof of the existence of a phase transition in all space dimensions. We have moreover calculated exactly the non-universal threshold value of the annihilation rate for this phase transition to occur. This result is beyond the possibilities of usual perturbation theory.

As for the parity conserving case, we find, surprisingly, that the appearance of the PC fixed point associated with an active-to-absorbing transition below a critical dimension must occur at a nonzero value of the branching rate, which would be compatible with a scenario where there are not one but two new fixed points as $d$ is lowered.

\begin{acknowledgments}
 The authors want to thank H. Chat\'e and B. Delamotte for a careful reading of the manuscript. We acknowledge partial support from the PEDECIBA program and ANII (Grant FCE-2009-2694).
\end{acknowledgments}

%%%%%%%%%%%%%%%%%%%%%%%%%%%%%%%%%
%%%%%%%%%%%%%%%%%%%%%%%%%%%%%%%%%
%\bibliography{mybib}

\begin{thebibliography}{99}
\bibitem{notepcpd}
BARW give also access to the Pair Contact Processes with Diffusion (PCPD) universality class \cite{chate,pcpd}, but we will not be dealing with it in this work.
\end{thebibliography}

%\bibliographystyle{plain}

\end{document}